\documentclass[twocolumn,secnumarabic,amssymb, nobibnotes, aps, prd]{revtex4-1}

\usepackage{color}
\usepackage{cancel}
\usepackage{ulem}
\usepackage{graphicx}
\usepackage{dcolumn}
\usepackage{bm}
\usepackage{textcomp}
\usepackage{amsmath}

\begin{document}
\title[]{   Magnetohydrodynamic  instabilities in a self-gravitating  rotating cosmic plasma}
\author{Jyoti Turi}
\email{ jyotituri.maths@gmail.com}
\affiliation{Department of Mathematics, Siksha Bhavana, Visva-Bharati (A Central University), Santiniketan-731 235, West Bengal, India}
\author{A. P. Misra}
\email{apmisra@visva-bharati.ac.in}
\affiliation{Department of Mathematics, Siksha Bhavana, Visva-Bharati (A Central University), Santiniketan-731 235, West Bengal, India} 
\date{\today}

\begin{abstract}
 The generation of magnetohydrodynamic (MHD) waves and their instabilities are studied   in  galactic gaseous rotating plasmas  with the effects of the magnetic field,  the self gravity, the diffusion-convection of cosmic rays as well as the  gas and cosmic-ray pressures. The coupling of the  Jeans, Alfv{\'e}n  and  magnetosonic waves, and the  conditions of  damping or instability are studied in three different cases, namely when the propagation direction is perpendicular, parallel and oblique to the static magnetic field, and are shown to be significantly modified by the effects of the Coriolis force due to the rotation of cosmic fluids and the cosmic-ray diffusion. The coupled  modes can be damped or anti-damped depending on the wave number is above or below the Jeans critical wave number that is reduced by the effects of the Coriolis force  and the cosmic-ray pressure.  It is   found that the deviation of the  axis of rotation from the direction of the static magnetic field   gives rise to the coupling between the Alfv{\'e}n wave and the classical Jeans mode which otherwise results into the modified slow and fast Alfv{\'e}n waves as well as the modified classical Jeans modes. Furthermore, due to the effects of the cosmic rays diffusion, there appears a  new wave mode (may be called the fast Jeans mode) in the intermediate frequency regimes of the slow and fast Alfv{\'e}n waves, which seems to be dispersionless in the long-wavelength propagation and has a lower growth rate  of instability in the high density regimes of galaxies.  The dispersion properties and the instabilities of different kinds of MHD waves reported here can play pivotal roles in the formation of various galactic structures at
different length scales.
\end{abstract}
\maketitle

\section{Introduction  } \label{sec-intro}
 One of the paramount examples of the magnetohydrodynamic (MHD) instability is the Parker instability \cite{parker1966}, which   has been  known to be relevant to the galaxy evolution, i.e., possibly the formation of molecular clouds and the galactic dynamo. Such instabilities are typically characterized by the length-scales that can be  shorter than the galactic radius. However, on relatively large scales, the Parker instability together with the Jeans instability \cite{jeans1902stability}, associated with the self-gravitational perturbations, can induce the formation of filament-like structures in the interstellar medium (ISM).  It has been shown that such instability produces most likely small-scale structures instead of the  giant molecular clouds \cite{asseo1978}.  Furthermore, the Parker instability may be significantly suppressed by means of fluctuating galactic magnetic fields \cite{kim2001} and so, other MHD instabilities should intervene to the formation of large-scale galactic structures. In this context,  the generation of magnetosonic waves and the formation of structures in galactic gaseous plasmas has been studied by Bonanno \textit{et al.} taking into account the effects of the differential rotation, the magnetic field and the self-gravity \cite{bonanno2008}. Recent investigations indicate that the modification of the Parker instability with the effects of cosmic-ray  transport can have large implications for the evolution and structure formation of galaxies \cite{heintz2020}. 
 \par 
On the other hand, the  magnetogravitational instability \cite{howard1962hydrodynamic} of space and astrophysical plasmas has been extensively studied   with the effects of the Earth's rotation, the self-gravitational force and the magnetic field \cite{gliddon1966gravitational,hamabata1984propagation,hamabata1985instability,hamabata1985propagation,singh1986gravitational,bora1991gravitational}.   
 To mention a few of them,   Prajapati \textit{et al.} \cite{prajapati2008self} have investigated the  self-gravitational instability in   anisotropic rotating plasmas and showed that the effects of  rotation can stabilize  the system.   Kossacki \cite{kossacki1961magnetogravitational} studied the Jeans instability in  viscous plasmas with the effects of the Coriolis force due to the rotation of fluids  and finite electric conductivity. In other  investigations,  many authors  have considered  the MHD instabilities in anisotropic   plasmas with the influences of the Hall current and finite electric and thermal conductivities  \cite{prajapati2008,chhajlani1985effect,kalra1970effect,ariel1970gravitational,vyas1988magnetogravitational,chhajlani1993stability,bhatia1995gravitational}. Ren \textit{et. al.} \cite{ren2009jeans} have advanced the theory of MHD waves in quantum magnetoplasmas by considering  the effects of  plasma resistivity.  The propagation  of MHD waves was also explored in self-gravitating dusty plasmas \cite{jacobs2004,sharma1982gravitational,chhajlani1986finite,chhajlani1994magnetogravitational}. The Jeans instability criteria in  two-component plasmas containing  ionized and neutral componants has also been discussed in other related works \cite{herrnegger1972effects,bhatia1973finite,chhonkar1977larmor,vaghela1990gravitational}. Recently,   
  Sharma  \textit{et al.} \cite{sharma2021modes} have studied the MHD instabilities   in finitely conducting neutrino-coupled magnetized plasmas by the effects of the self-gravity and dissipation.
  \par The aim of this work is to advance the theory of MHD waves   and their instabilities in ISM consisting of thermal gas and cosmic rays using a two-fluid approximation. Especially, we focus on the generation of different kinds of wave modes including new type of coupling and the generation of a new wave mode, not reported before,  as well as the characteristics of wave damping and instabilities that are modified by the Coriolis force due to the rotation of cosmic fluids, the self-gravitational force, the magnetic field, the thermal gas and cosmic-ray pressures as well as the cosmic-ray diffusion.  The paper is organized as follows: In Sec. \ref{sec-basic-eq}, we describe the two-fluid MHD model for an ionized thermal gas and cosmic rays, and obtain a general dispersion relation for MHD waves. Section \ref{sec-disp} demonstrates the  characteristics of wave dispersion and damping/instability in three different cases of propagation (perpendicular, parallel and oblique to the magnetic field). Finally, the results are summarized and concluded in Sec. \ref{sec-summ}.  
\section{Basic equations} \label{sec-basic-eq}
We consider the propagation of MHD waves in an interstellar   medium  that consists of a thermal ionized gas and cosmic rays under the influences of a self-gravitational force, the magnetic force, the pressure gradient force, and  the Coriolis force due to  the rotation of cosmic fluids with an angular velocity $\mathbf{\Omega}=(0,\Omega_0 \cos\lambda,\Omega_0 \sin\lambda)$, where $\lambda$ is the angle made by the  axis of rotation with the $y$-axis. Here, the cosmic rays are considered to be a   gas with a negligible density but a significant contribution to the pressure \cite{bonanno2008}. We consider a polytropic thermal gas, i.e.,   no diffusion along any direction but only the compression and rarefaction processes, which include heat transfer, can take place, and  describe the evolution of cosmic rays in the diffusion approximation along the magnetic field lines  as they  are strongly magnetized and the diffusion transverse to the magnetic field is thus less important \cite{giacalone1999}.   The plasma is assumed to be   homogeneous,   fully ionized, and highly conducting  with a higher Reynolds number.  Furthermore, we assume that the uniform magnetic field is along the $z$-axis, i.e.,  $\mathbf{B_0}=B_0\hat{z}$ and ignore the effect of the centrifugal force compared to the Coriolis force.    The basic equations governing the dynamics of    MHD  waves are
\begin{equation}
	\label{eq-cont}
	\frac{\partial \rho}{\partial t} +\mathbf{\nabla}\cdot\left(\rho \mathbf{v}\right)=0, 
\end{equation} 
\begin{multline}
	\label{eq-moment}
	\frac{\partial \mathbf{v}}{\partial t} +\left(\mathbf{v}\cdot\mathbf{\nabla}\right) \mathbf{v}=-\frac{1}{\rho}\mathbf{\nabla}\left(P+\frac{\mathbf{B}^2}{2\mu_0}\right)\\
	+\frac{1}{\rho \mu_0} \left(\mathbf{B}\cdot\mathbf{\nabla}\right)\mathbf{B}-2\mathbf{\Omega}\times\mathbf{v} +\mathbf{\nabla}\psi,
\end{multline}

\begin{equation}
	\label{eq-B}
	\frac{\partial \mathbf{B}}{\partial t} +\left(\mathbf{v}\cdot\mathbf{\nabla}\right) \mathbf{B}=\left(\mathbf{\textbf{B}}\cdot\mathbf{\nabla}\right) \mathbf{v}-\left(\mathbf{\nabla}\cdot\mathbf{v}\right) \mathbf{B},  
\end{equation}
\begin{equation}
	\label{eq-psi}
	\mathbf{\nabla}^2 \psi=-4\pi G \rho,
\end{equation}
together with the equations of state for the thermal gas and the diffusion-convection equation for cosmic rays \cite{bonanno2008}:
\begin{equation}\label{eq-Pg}
\frac{\partial P_g}{\partial t} +\left(\mathbf{v}\cdot\mathbf{\nabla}\right) P_g+\gamma_g P_g \nabla\cdot\mathbf{v}=0,
\end{equation}
\begin{equation}\label{eq-Pc}
\frac{\partial P_c}{\partial t} +\left(\mathbf{v}\cdot\mathbf{\nabla}\right) P_c+\gamma_c P_c \nabla\cdot\mathbf{v}=\kappa\nabla\cdot\left(\nabla_\parallel P_c\right).
\end{equation}
Here, $\rho$ and $\mathbf{v}$ are, respectively, the fluid (thermal gas) mass density and velocity; $P=P_g+P_c$ is the total pressure in which $P_g$ is the gas pressure and $P_c$ that of cosmic rays;  $\mu_0$ is the vacuum permeability, $\psi$ is the gravitational potential, $G$ is the gravitational constant, $\kappa$ is the diffusion coefficient of cosmic rays along the magnetic field, and   $ \nabla_\parallel P_c=\mathbf{B}\left(\mathbf{B}\cdot\nabla P_c\right)/B^2$. Also, $\gamma_g$ and $\gamma_c$ are the adiabatic indices corresponding to the thermal gas and cosmic rays.   In the following section \ref{sec-disp}, we will obtain a general linear  dispersion relation from these basic set of equations using a perturbative approach.    

\section{Dispersion relation} \label{sec-disp}
In equilibrium state, we assume that  the plasma is uniform with constant density $\rho_0$ and  zero velocity. To linearize Eqs. \eqref{eq-cont}-\eqref{eq-Pc}, we split up the physical quantities into their equilibrium (with suffix $0$) and perturbation (with suffix $1$) parts according to $\rho=\rho_{0}+\rho_{1}$, $\mathbf{v}=0+\mathbf{v_1}$, $\mathbf{B}=\mathbf{B_0}+\mathbf{B_1}$, $\psi=\psi_0 + \psi_1$,  $P_{j}=P_{j0}+P_{j1}$ (with $j=g,~c$), and $P=P_0+P_1$ with $P_0=P_{g0}+P_{c0}$ and $P_1=P_{g1}+P_{c1}$.  Thus, we obtain
\begin{equation}
	\label{eq-cont1}
	\frac{\partial \rho_1}{\partial t} +\rho_{0} \mathbf{\nabla}\cdot\mathbf{v_1}=0, 
\end{equation}
\begin{multline}
	\label{eq-moment1}
	\frac{\partial \mathbf{v_1}}{\partial t} =-\frac{1}{\rho_0}\mathbf{\nabla}P_{1} -\frac{1}{\mu_0\rho_0}\mathbf{\nabla}\left(\mathbf{B_0}\cdot\mathbf{B_1}\right)
	+\frac{1}{\rho_0 \mu_0} \left(\mathbf{B_0}\cdot\mathbf{\nabla}\right)\mathbf{B_1}\\
	-2\mathbf{\Omega}\times\mathbf{v_1} +\mathbf{\nabla}\psi_1,
\end{multline}
\begin{equation}
	\label{eq-B1}
	\frac{\partial \mathbf{B_1}}{\partial t}  =\left(\mathbf{\mathbf{B_0}}\cdot\mathbf{\nabla}\right) \mathbf{v_1}-\left(\mathbf{\nabla}\cdot\mathbf{v_1}\right) \mathbf{B_0}, 
\end{equation}
\begin{equation}
	\label{eq-psi1}
	\mathbf{\nabla}^2 \psi_1=-4\pi G \rho_1,
\end{equation}
\begin{equation}\label{eq-Pg1}
\frac{\partial P_{g1}}{\partial t} +\gamma_g P_{g0} \nabla\cdot\mathbf{v_1}=0,
\end{equation}
\begin{equation}\label{eq-Pc1}
\left(\frac{\partial }{\partial t}+\nu_{c}\right)P_{c1}  +\gamma_c P_{c0} \nabla\cdot\mathbf{v_1}=0,
\end{equation}
where the frequency (inverse time scale) of cosmic rays  diffusion $\nu_c$ is given by \cite{bonanno2008} $\nu_cP_{c1}=-\kappa\nabla\cdot\left[ \mathbf{B_0}\left(\mathbf{B_0\cdot\nabla}\right) P_{c1}/B_0^2\right]$. 
\par 
Next,   assuming the perturbations to  vary as plane waves of the form $\sim\exp(i \mathbf{k}\cdot \mathbf{r}-i \omega t)$ with the wave vector $\mathbf{k}$ and wave frequency $\omega$, we obtain    from Eqs. \eqref{eq-cont1} to \eqref{eq-Pc1}   the following general dispersion relation. 
\begin{multline}
	\label{eq-disp}
	 \omega^2  \mathbf{v_1} -\left(C_g^2+\frac{ C_c^2\omega}{\omega+i\nu_c} \right)\left(\mathbf{k}\cdot \mathbf{v_1}\right) \mathbf{k}\\
	 -\frac{\mathbf{k}}{\mu_0 \rho_{0}}\left[(\mathbf{k}\cdot \mathbf{v_1}) B_0^2-(\mathbf{B_0}\cdot \mathbf{k})(\mathbf{B_0}\cdot \mathbf{v_1})\right]\\
	-\frac{\mathbf{B_0}\cdot \mathbf{k}}{\mu_0 \rho_{0}}\left[(\mathbf{k}\cdot \mathbf{v_1}) \mathbf{B_0}-(\mathbf{B_0}\cdot \mathbf{k})\mathbf{v_1}\right]\\
	+i 2\omega \mathbf{\Omega} \times \mathbf{v_1} +\frac{4 \pi G\rho_{0}}{k^2}\left(\mathbf{k}\cdot \mathbf{v_1}\right) \mathbf{k}=0,
\end{multline}
where the reduced expression of $\nu_c$ is given by $\nu_c=\kappa\left(\mathbf{k}\cdot\mathbf{B_0}\right)^2/B_0^2$.  Also, $C_g=\sqrt{\gamma_g P_{g0}/\rho_0}$ is the sound speed and $C_c=\sqrt{\gamma_c P_{c0}/\rho_0}$ that associated with  cosmic rays. 
\par 
In what follows, we consider the  propagation of MHD waves at an arbitrary direction with respect to the static magnetic field $\mathbf{B_0}$. Without loss of generality, we assume that $\mathbf{k}=k\left(\hat{x}\sin\theta+\hat{z}\cos\theta\right)$, where $\theta$ is the angle between  $\mathbf{k}$  and $\mathbf{B_0}$.  Also,    $\mathbf{v_1}=(v_x,v_y,v_z)$  in which  $v_x$, $v_y$, and $v_z$ denoting the perturbed velocity components along  $x$, $y$, and $z$ axes. Thus,  Eq. \eqref{eq-disp}  reduces to
  \begin{multline}
	\label{eq-obliq}
	\left( \omega^{2}-\tilde{C_s^{2}} k^{2}\cos^2\theta +\omega_{J}^{2}\cos^2\theta\right) \\
	\times \left[\left\lbrace\omega^{2}-k^{2}V_A^2+\left(\omega_J^2-\tilde{C_s^{2}}k^2\right)\sin^2\theta\right\rbrace\left(\omega^{2}-k^{2}V_A^2\cos^2\theta\right)\right.\\
\left.	-4\Omega_0^2 \omega^{2} \sin^2\lambda\right]-\left(\omega^{2}-k^{2}V_A^2\cos^2\theta\right) \\
\times	\left[4 \Omega_0^2 \omega^{2}\cos^2\lambda  + \left(\omega_J^2-\tilde{C_s^{2}}k^2\right)^2\sin^2\theta\cos^2\theta\right]=0,
\end{multline} 
where   $\tilde{C_s^{2}}=C_g^2+ {C_c^2 \omega}/\left(\omega+i\nu_c\right)$ with $C_s^{2}=C_g^2+ C_c^2$ denoting the squared  acoustic speed in absence of the cosmic rays diffusion $(\nu_c=0)$, $\omega_J=\sqrt{4 \pi \rho_0 G}$ is the classical Jeans frequency and $V_A=B_0/\sqrt{\rho_0\mu_0}$ is the Alfv{\'e}n velocity.
\par
  The dispersion equation \eqref{eq-obliq} is, in general, complex due to the appearance of the factor $\omega+i\nu_c$. It can be analyzed for the identification of different kinds of MHD wave modes, their propagation characteristics as well as their stability with the effects of the self-gravitational force, the Coriolis force, and   the dissipation due to the cosmic rays diffusion. Furthermore, Eq. \eqref{eq-obliq}  can be studied  in  three different  cases of interest, namely when (i) the propagation vector is perpendicular to the magnetic field, i.e.,  $\mathbf{k}\perp \mathbf{B_0}$ with $\theta=\pi/2$, (ii)  the propagation vector is parallel to the magnetic field, i.e.,  $\mathbf{k}\parallel \mathbf{B_0}$ with $\theta=0$, and (iii)   the propagation direction is arbitrary, i.e., the  angle $\theta$ assumes any values in $0\leq\theta\leq\pi/2$.   
  \par 
It is to be mentioned that   for the sake of simplicity, we have neglected the effects of particle collision by assuming that the collision time is much longer than the time scale of oscillations. Also, in a highly conducting medium with high Reynolds number, both the effects of fluid viscosity and magnetic viscosity (plasma registivity) on the attenuation of MHD waves can be shown to be  small (Although the attenuation increases with the   viscosity effects, the same  decreases as the fluid conductivity increases).  On the other hand, it has been  shown that for large-scale structure formation, the  two possible sources of dissipation due to self-interaction and gravitational coupling of cosmic fluids  can operate together and  the interplay between them can play an important role in determining the dynamics of   cosmic fluids \cite{natwariya2020}. However,  inclusion of such dissipative effects  is beyond the scope of the present study.   
\subsection{Propagation perpendicular to the magnetic field} \label{sec-sub1}
We consider the direction of propagation across the  magnetic field, i.e.,  $\mathbf{k}=k \hat{x}$, $\mathbf{B_0}=B_0 \hat{z}$ with $\theta=\pi/2$ and $\mathbf{\Omega}=(0,\Omega_0\cos\lambda,\Omega_0\sin\lambda)$. In this case, the  general dispersion relation  \eqref{eq-obliq} reduces to
 \begin{equation}
	\label{eq-disp1}
	\omega^2=\left(\tilde{C_s^{2}} +V_A^2\right)k^2-\omega_J^2 +4\Omega_0^2, 
\end{equation}
  Equation \eqref{eq-disp1} describes  the wave dispersion and instability for the Jeans-Alfv{\'e}n-magnetosonic (JAM) wave  which is influenced by the thermal pressure due to gas $(C_g)$ of charged particles and cosmic rays $(C_c)$, the cosmic rays diffusion $(\nu_c)$,  the external magnetic field $(V_A)$, the gravitational force $(\omega_J)$, and the Coriolis force  $(\Omega_0)$ due to the  rotation of cosmic fluids.   By disregarding the effects of the cosmic rays pressure and diffusion,   the Coriolis force and the external magnetic field, one can recover the classical Jeans mode in electron-ion  plasmas, given by,
\begin{equation}
\omega^2=C_g^{2}k^2-\omega_J^2,
\end{equation}
 which is known to be stable (unstable) for $k>k_{J_c}~(k<k_{J_c})$, where $k_{J_c}=\omega_J/C_g$ is the Jeans critical wave number. Also, in absence of the gas and cosmic rays pressures (or if the  contributions of these pressures are much smaller than that of the magnetic pressure)  together with the the gravitational force and the Coriolis force, one obtains the pure Alfv{\'e}n wave with the phase velocity $\omega/k=V_A$. Furthermore, the typical magnetosonic mode can be recovered with the phase velocity $\omega/k=\sqrt{C_g^2+V_A^2}$ if one neglects the cosmic rays pressure (or when the gas pressure dominates over the cosmic rays pressure),  the gravitational force and the Coriolis force.  Such  magnetosonic (or magnetoacoustic) waves  become  ion-acoustic waves with velocity $C_g$ if the gas pressure is much higher than the  cosmic rays pressure and the magnetic pressure. Thus, it follows that in presence of all these aforementioned effects,   a new coupled JAM mode is  generated which can be damped or anti-damped due to the effects of the cosmic rays diffusion and the interplay between the Coriolis force and the self-gravity. 
\par  We note that  Eq. \eqref{eq-disp1} is also, in general, complex  due to the effect of the   cosmic rays diffusion $(\nu_c)$. Even in absence of this effect,   the Jeans instability criterion for the JAM mode is significantly modified by the  cosmic rays pressure and the Coriolis force.    In the case of $\nu_c=0$,    the JAM  mode is always stable for $\omega_J\lesssim2\Omega_0$ or when the influence of the self-gravitational force is negligible (i.e., $\omega_J=0$). However, for $\omega_J>2\Omega_0$, the mode can be stable or unstable according to when $k>k_{J_1}$ or $k<k_{J_1}$, where $k_{J_1}$ is the modified Jeans critical wave number, given by,
\begin{equation} \label{eq-kJ1}
k_{J_1}=\left(\frac{\omega_J^2-4\Omega_0^2}{C_s^2+V_A^2} \right)^{1/2}.
\end{equation}
  Thus, the Jeans instability criterion is not only  modified, but the critical wave number is also significantly reduced due to the cosmic rays pressure and the Coriolis force. It follows that the JAM mode or the gravitational mode becomes stable (unstable) in the longer (shorter) domains of   wavelengths   than that predicted for the classical Jeans mode \cite{jeans1902stability}.  This means that the effects of  rotation and the cosmic rays pressure favor the stability of JAM modes in cosmic plasmas.  Typically,  for plasmas relevant to spiral galaxies \cite{gliddon1966gravitational} we have $B_0\sim1$ nT, $P_{g0},~P_{c0}\sim10^{-13}$ Nm$^{-2}$, $\rho_0\sim10^{-21}$ Kg/m$^3$. So, considering  $\omega_J\sim0.5\omega_0$, $\Omega_0 \sim0.2\omega_0$ and $V_A/C_s\sim0.5$, where $\tau\equiv\omega_0^{-1}$ is the typical time scale of oscillations ($\sim10^{15}$ s),  the critical wavelength can be estimated as $k_{J_1}^{-1}\sim3.7C_s/\omega_0$, and the maximum instability growth rate as $\gamma_\text{max}\sim\sqrt{\omega_J^2-4\Omega_0^2}\sim0.3\omega_0$ at $k=0$. The corresponding growth time of instability can  then be estimated as $\tau_\text{growth}\sim10^8$ yrs, which is comparable to the evolutionary time of diffuse interstellar clouds. 
\par
\begin{figure*}
	\includegraphics[width=6.5in,height=2.5in]{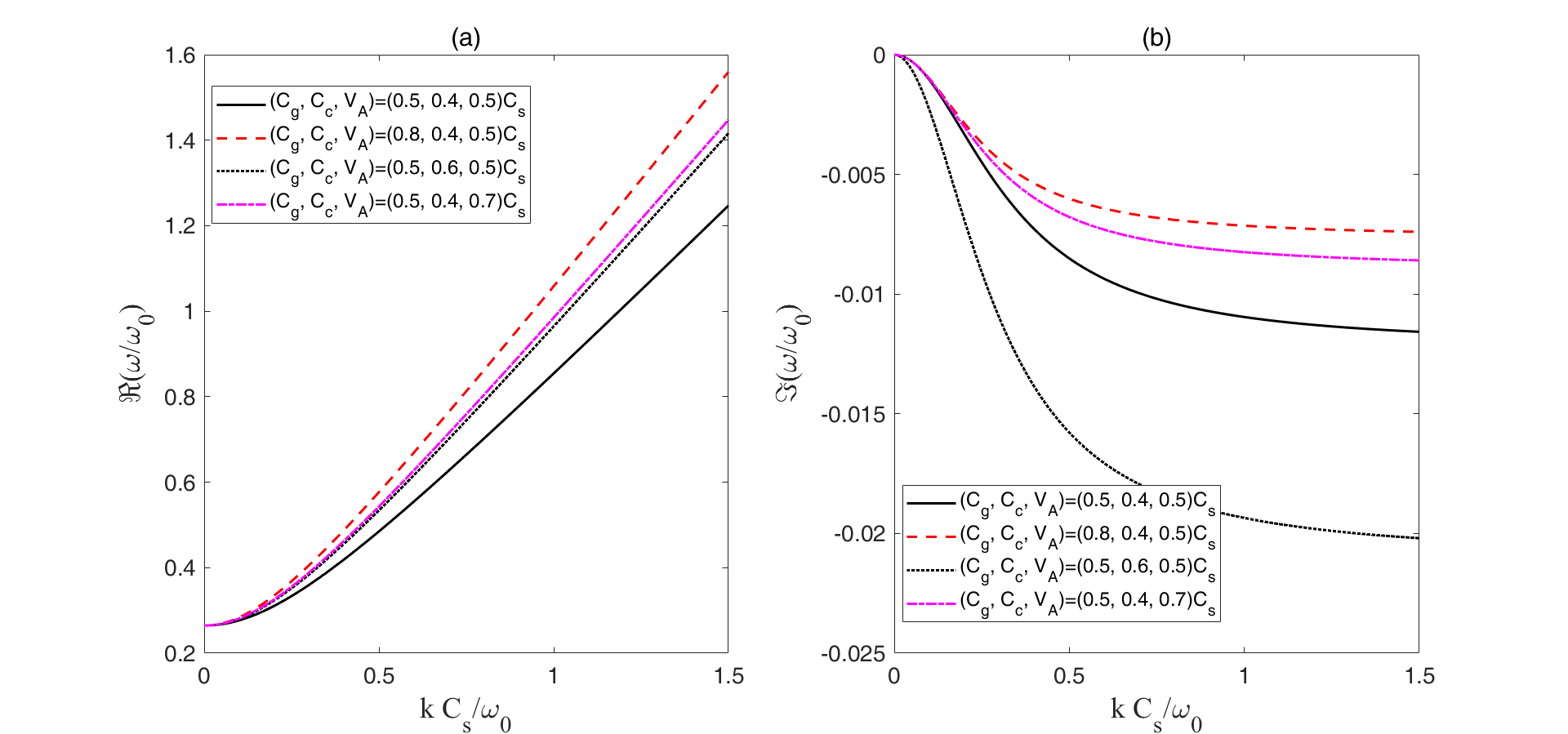}
	\includegraphics[width=6.5in,height=2.5in]{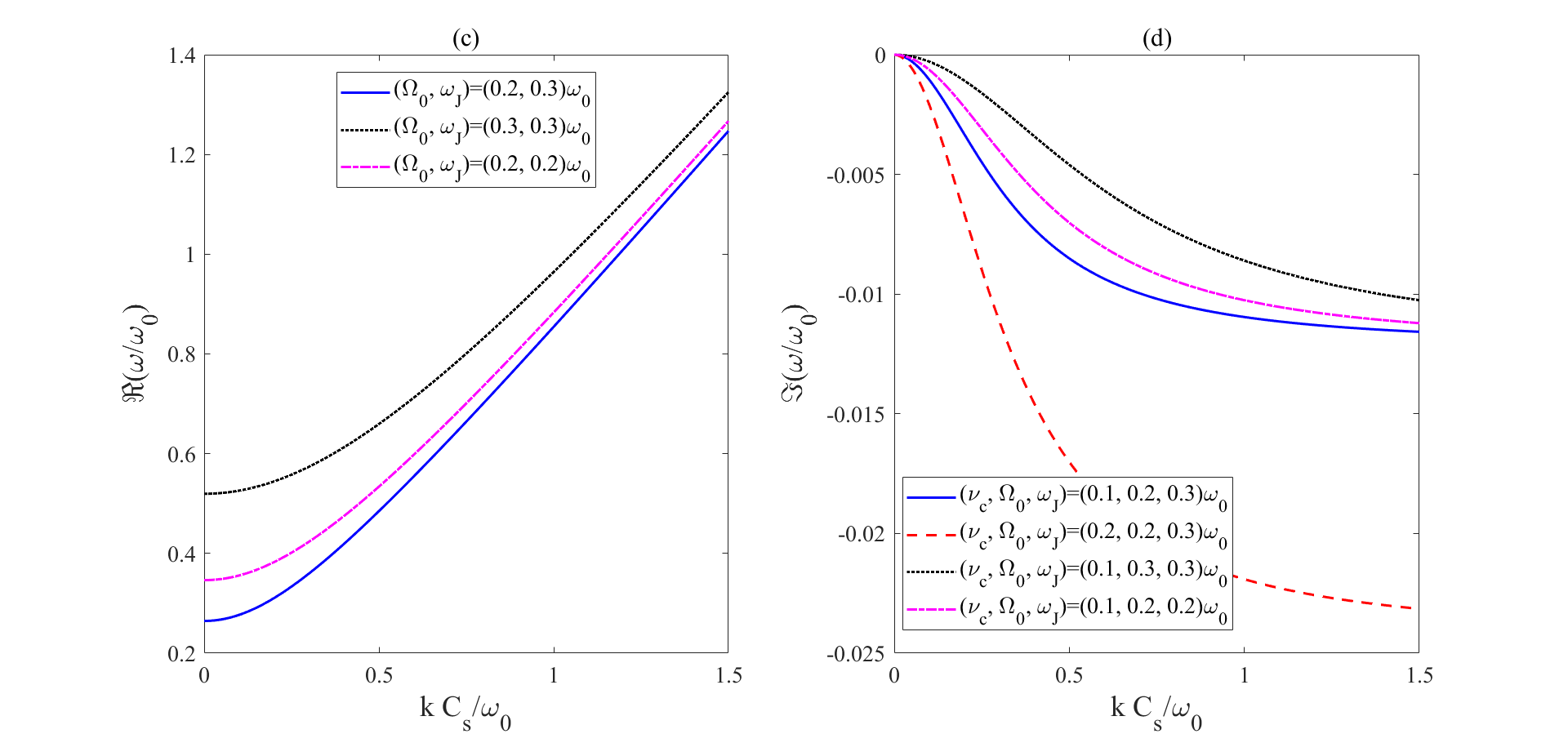}
	\caption{The dispersion properties [subplots (a) and (c)] and the damping rates [subplots (b) and (d)] are shown for the Jean-Alfv{\'e}n-magnetosonic wave [Eq. \eqref{eq-disp1}] in the case of $\omega_J\lesssim2\Omega_0$. The fixed parameter values for subplots [(a), (b)] and [(c), (d)], respectively, are $(\nu_c,~\Omega_0,~\omega_J)=(0.1,~0.2,~0.3)\omega_{0}$  and $(C_g,~C_c,~V_A)=(0.5,~0.4,~0.5)C_s$.        }
	\label{fig1}
\end{figure*}

\begin{figure*}
	\includegraphics[width=6.5in,height=3in]{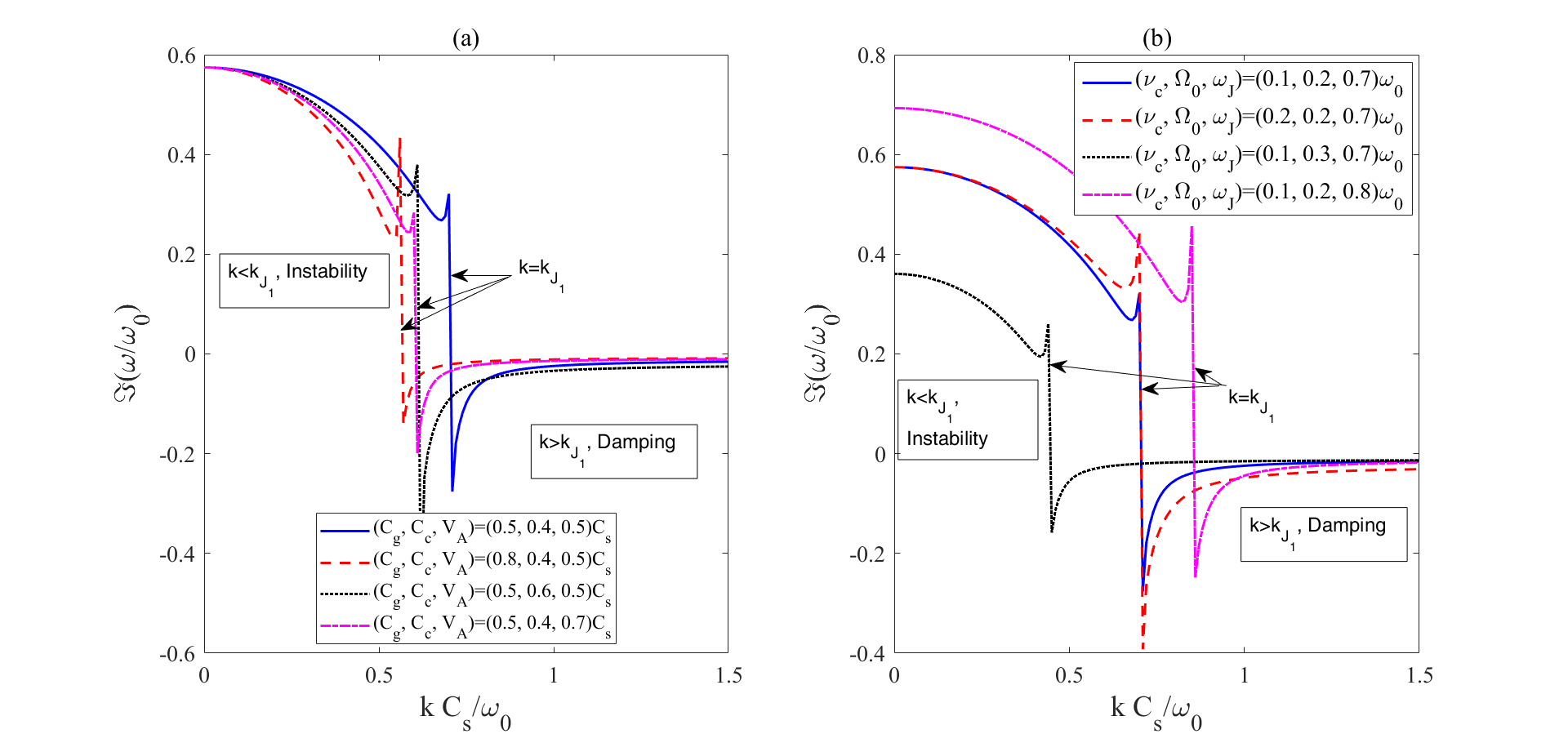}
		\caption{The   growth and damping rates  are shown for the Jean-Alfv{\'e}n-magnetosonic wave [Eq. \eqref{eq-disp1}] in the case of $\omega_J>2\Omega_0$. The fixed parameter values for subplots [(a), (c)] and [(b), (d)] are the same as in Fig. \ref{fig1}.   The peaks of the curves appear at the critical value $k=k_{J_1}$.  The wave instability occurs $\left[\Im{(\omega/\omega_{0})}>0\right]$ in the region $k<k_{J_1}$ and the wave gets damped $\left[\Im{(\omega/\omega_{0})}<0\right]$ in the other region $k>k_{J_1}$.  }
	\label{fig2}
\end{figure*}

  On the other hand, in presence of the cosmic rays diffusion, the JAM mode becomes unstable, i.e.,  it can be either damped or anti-damped. The expressions for the damping and growth rates can be obtained from the dispersion equation \eqref{eq-disp1} by assuming $\omega=\omega_r+i\gamma$;   $|\nu_c|,~ |\gamma|\ll\omega_r$;  $\omega/(\omega+i\nu_c)\approx 1–i\nu_c/\omega_1$, and using the conditions $\omega_J\lesssim \Omega_0$ and $\omega_J> \Omega_0$ separately.  In the former case, $\omega_1$ is real and $\omega_r=\omega_1$, while in the latter, $\omega_1$ is real $(=\omega_r)$ or purely imaginary according to when $k>k_{J_1}$ or $k<k_{J_1}$. Here, $\omega_1$ is a solution of Eq. \eqref{eq-disp1}  at $\nu_c=0$. Thus,   when $\omega_J\lesssim2\Omega_0$, the JAM  wave gets damped and the damping rate is   $|\gamma|\sim(1/2)C_c^2k^2\nu_c/\omega^2$, where $\omega$ is the   the wave frequency at $\nu_c=0$.  However, when  $\omega_J>2\Omega_0$,    the  instability (damping) occurs for $k<k_{J_1}~(k>k_{J_1})$. In this case,  the 
damping rate has  the same expression as above, however, the growth rate of instability can be estimated as $\gamma\sim \left(\omega^2+C_c^2k^2\nu_c/\omega\right)^{1/2}$ with $\omega$ denoting the wave frequency at $\nu_c=0$. In order to study the characteristics of the JAM mode in details we numerically solve Eq. \eqref{eq-disp1}. The results are displayed in Figs. \ref{fig1} and \ref{fig2} corresponding to the cases with   $\omega_J\lesssim2\Omega_0$   and $\omega_J>2\Omega_0$  respectively. In the former, while only the damping takes place irrespective of the values of $k$, in the latter,   the damping or instability (anti-damping) can occur according to when $k>k_{J_1}$ or $k<k_{J_1}$. 
\par
 From Fig. \ref{fig1} it is seen that  the real part of the wave frequency increases with $k$. Also,  the absolute value of the damping rate  increases within the domain $0\leq k\lesssim1$ and it reaches a steady state value for $k\gtrsim1$.      This means that the wavelength is decreased to accommodate a large  number of wave modes to pass through a given point   per unit time. However, they die out within a short interval of time due to the higher damping rate.    From the subplots (a) and (b)     it is noted that while the  wave frequency (real part, $\Re{\omega}$) is increased,  the damping rate (imaginary part, $|\Im\omega|$) is significantly reduced due to a small increment of the  parameters $C_g$ and $V_A$ associated with the thermal gas pressure and the static magnetic field (Compare the solid line with the dashed and dash-dotted lines).    A significant increase of both the wave frequency  and the damping rate also occurs  by the effects of the cosmic ray pressure ($C_c$)  (Compare the solid lines with the dotted lines).    On the other hand, subplots (c) and (d) of Fig. \ref{fig1} show that the effects of the rotational frequency $(\Omega_0)$ with its increased value and the Jeans frequency $(\omega_J)$ with a small reduction are to enhance the wave frequency but to reduce the damping rate significantly (Compare the solid line with the   dotted and dash-dotted lines).  As expected, a small increase of the diffusion frequency $\nu_c$ leads to a significant enhancement of the damping rate [See the dashed lines in subplot (d)]. Thus, it may be concluded that for the propagation of JAM waves in rotating magnetoplasmas, the damping rate can be diminished by  either increasing   the thermal gas pressure/ the magnetic field strength   or decreasing the  cosmic  rays pressure/ Jeans frequency. So,   highly magnetized gaseous plasmas  with  dominating effects from the Coriolis force over the self-gravity can exhibit the stabilizing behaviors of JAM modes.
\par 
 Figure  \ref{fig2} shows the  plots of the growth and the damping rates in the case when the contribution from the self-gravity is higher than that due to the Coriolis force, i.e., $\omega_J>2\Omega_0$. As   discussed before, the instability (damping) occurs for $k<k_{J_1}$ $(k>k_{J_1})$. The peaks of the curves appear at the critical value, $k=k_{J_1}$ at which the damping or the growth rate of the JAM mode may  not be defined.  However, the growth (decay) rate tends to decrease with increasing values of $k$ in the interval $0\lesssim k<k_{J_1}$ $(k>k_{J_1})$. The growth rate becomes minimum or the damping rate gets maximized close to  this critical value.  Here, for larger values of  $k~(>k_{J_1})$, the damping rate  reaches a steady state. Furthermore, due to a small increment of the parameters $C_g$, $C_c$ and $V_A$, associated with the gas pressure, cosmic rays pressure and the static magnetic field, the growth or damping rates are reduced together with a shift of the critical value 
$k_{J_1}$   towards the lower region of $k$, i.e, the domain of $k$ for the wave instability shortens while that for the damping is extended  [See subplot (a)].  On the other hand, the effect of an increment of the   diffusion frequency $\nu_c$ is to enhance both the growth and the damping rates even though the critical wave number $k_{J_1}$  remains the same. However, in contrast to the effects of the rotational frequency (with a small increase of which both the growth and the damping rates are significantly reduced together with a shift of the critical wave number towards lower values of $k$), the effects of the Jeans frequency are to enhance both the growth and the damping rates as well as an increase of the critical wave number. The latter  results into an increase of the instability domain $0<k<k_{J_11}$ but a decrease of the domain for damping $(k>k_{J_1})$. So, it may be concluded that when the self-gravitational force dominates over the Coriolis force due to the  rotation of fluids, cosmic magnetoplasmas   always exhibit instability. Such magneto-Jeans instability can have important contributions to the  formation of large-scale clouds   in spiral arms or galactic centers \cite{kim2002}.
\subsection{Propagation parallel to the magnetic field  } \label{sec-sub2}
We consider the  propagation of MHD waves parallel to the static magnetic field, i.e., $\mathbf{k}=k \hat{z}$,  $\mathbf{B_0}=B_0 \hat{z}$, $\theta=0$, and $\mathbf{\Omega}=(0,\Omega_0 \cos\lambda,\Omega_0 \sin\lambda)$. In this case, the dispersion relation \eqref{eq-obliq}
reduces to
\begin{multline}
	\label{eq-disp2}
	\left( \omega^{2}-\tilde{C_s^{2}} k^{2} +\omega_{J}^{2}\right)\left[ \left(\omega^{2}-k^{2}V_A^2\right)^2-4\Omega_0^2 \omega^{2} \sin^2\lambda\right]\\
	-4 \Omega_0^2 \omega^{2}\cos^2\lambda  \left(\omega^{2}-k^{2}V_A^2\right)=0,
\end{multline} 
where $\tilde{C_s^{2}}=C_s^2$ at $\nu_c=0$.
Equation \eqref{eq-disp2} represents the dispersion relation for coupled Jeans and Alfv{\'e}n wave modes, to be called Jeans-Alfv{\'e}n (JA) modes that are typically  modified by the cosmic-ray pressure and diffusion, as well as the Coriolis force.  Interestingly, the coupling of these modes occurs due to the obliqueness of the  axis of rotation with respect to the direction of the magnetic field and they  become decoupled when $\mathbf{\Omega}\parallel \mathbf{B_0}$ or $\lambda=\pi/2$.   We note that Eq. \eqref{eq-disp2} is, in general, complex, which can exhibit either the wave instability or damping depending on the system parameters as stated before and the wave number below or above a critical value.  Before we proceed to the general case, we first examine the nature of the roots of Eq. \eqref{eq-disp2} in absence of the cosmic-ray diffusion $(\nu_c=0)$. To this end, we recast Eq. \eqref{eq-disp2}   as a polynomial equation in $\omega^2$, i.e., 
\begin{equation}
\begin{split}
	\label{eq-disp3}
	\left(\omega^2\right)^3&+\left(\omega_J^2- C_s^2k^2-2V_A^2k^2-4\Omega_0^2\right)(\omega^2)^2\\
&-\left[\left(\omega_J^2-C_s^2k^2\right)\left(2V_A^2k^2+4\Omega_0^2 \sin^2\lambda\right)\right.\\
	&-\left.\left(V_A^4k^4+4\Omega_0^2 \cos^2\lambda k^2V_A^2\right)\right]\omega^2 \\
	&+\left(\omega_J^2-C_s^2k^2\right)V_A^4k^4=0.
	\end{split}
\end{equation}
  When the contribution of the self-gravitational force is higher than that of the Coriolis force, i.e.,  $\omega_J>2\Omega_0$, applying the Decartes'   rule of signs,   we note that  Eq. \eqref{eq-disp3} has a maximum of two positive roots  (i.e., a maximum of two stable wave modes)  and a maximum  of one negative root  (i.e.,  either one unstable mode or no unstable mode) for  $k<k_{J_2}$  and   $k>k_{J_2}$, where $k_{J_2}~(<k_{J_1})$ is the reduced critical wave number, given by,
  \begin{equation} \label{eq-kJ2}
k_{J_2}=\left(\frac{\omega_J^2-4\Omega_0^2}{C_s^2+2V_A^2} \right)^{1/2}.
\end{equation}  
Similarly, considering   $\omega_J\leq2\Omega_0$, one can also conclude that there may be two stable wave modes and one unstable mode (or no unstable mode) for  $k<k_{J_3}$  and   $k>k_{J_3}$ respectively, where $k_{J_3}\equiv\omega_J/\sqrt{C_s^2+V_A^2}<k_J\equiv \omega_J/C_s$.  
\par 
 In particular, for  $\lambda=\pi/2$, i.e., when the  axis of rotation is along the static magnetic field $(\mathbf{\Omega}\parallel\mathbf{B_0})$ for which   the  Jeans mode and  the   Alfv{\'e}n mode get decoupled, one obtains from Eq. \eqref{eq-disp3} the following dispersion relation for the classical Jeans mode but modified by the cosmic-ray pressure. 
\begin{equation}
		\omega^2-C_s^2 k^2 +\omega_J^2=0. \label{eq-J1}
\end{equation}
This gives an unstable (or a stable) mode for $k<k_J$ ($k>k_J$). In this case, we  obtain the dispersion relations for   
  the   fast  and  slow  Alfv{\'e}n waves  modified by the Coriolis force, i.e., 
\begin{equation}\label{eq-alfven}
	\omega=\pm\Omega_0+\sqrt{\Omega_0^2+k^2 V_A^2}.
\end{equation}
While for the slow mode  the cut-off frequency is zero,  for the fast mode it is shifted by $\omega=2\Omega_0$. Both the modes are clearly stable being independent of the self-gravity effect. 
\par 
On the other hand, when $\lambda=0$, i.e., when the  axis of rotation is transverse to the direction of the magnetic field,  Eq. \eqref{eq-disp3} reduces to 
\begin{multline}
	\label{eq-disp-lamb0}
	\left(\omega^2-k^2V_A^2\right)\left[\left(\omega^2-C_s^2k^2+\omega_J^2\right)\left(\omega^2-k^2V_A^2\right)\right.\\
	\left.-4\Omega_0^2\omega^2\right]=0.
\end{multline} 
The first factor of Eq. \eqref{eq-disp-lamb0}, when equated to zero, gives the pure Alfv{\'e}n mode having the phase velocity as the Alfv{\'e}n velocity $V_A$. The second factor gives the following reduced dispersion relation for the coupled Jeans and the Alfv{\'e}n modes.
 \begin{equation}
	\label{eq-disp-lamb01}
	\left(\omega^2-C_s^2+\omega_{J}^2\right)\left(\omega^2-k^2V_A^2\right)-4\Omega_0^2\omega^2=0.
\end{equation}
For further insights of these modes, we express Eq. \eqref{eq-disp-lamb01}   as a quadratic equation in $\omega^2$, i.e., 
\begin{equation}
	\label{eq-disp-lamb02}
	\left(\omega^2\right)^2+\left[\omega_{J}^2-\left(C_s^2+V_{A}^2\right)k^2-4 \Omega_0^2\right] \omega^2-\omega_{J}^2 V_{A}^2 k^2=0.
\end{equation}
Here, two particular cases  may be of interest, namely when   $\omega_{J}>2\Omega_0$ and $\omega_{J}\lesssim2\Omega_0$. In the former case, the dispersion relation  \eqref{eq-disp-lamb02}
is rewritten as
\begin{equation}
	\label{eq-disp-lamb03}
	\left(\omega^2\right)^2+\left(\omega_{J0}^2-C_S^2 k^2-V_{A}^2k^2\right) \omega^2-\omega_{J}^2 V_{A}^2 k^2=0,
\end{equation}
where $\omega_{J0}^2=|\omega_{J}^2-4\Omega_0^2|$ is the square of the reduced Jeans frequency. Solving Eq. \eqref{eq-disp-lamb03} for $\omega^2$,    we obtain   two different roots of $\omega^2$, given by,

\begin{multline}
	\label{eq-disp-lamb04}
	\omega_{1,2}^2=\frac{1}{2}\left[k^2\left(C_s^2+V_A^2\right)-\omega_{J0}^2\right.  \\
	 \left. \pm\sqrt{\left[\omega_{J0}^2-k^2\left(C_S^2+V_A^2\right)\right]^2+4 k^2 V_A^2\omega_{J}^2}\right].	
\end{multline}
 We note that the expression for $\omega_1^2$ is always positive, giving a real root, while that of $\omega_2^2$ is always negative (giving an imaginary root) irrespective of the values of $k$. Thus, it follows that  the fast (slow) Jeans-Alfv{\'e}n mode is always stable (unstable) for $\omega_J>2\Omega_0$. The instability growth rate for the slow mode is 
\begin{multline}
	\gamma=\frac{1}{\sqrt{2}}\left[\sqrt{\left[\omega_{J0}^2-k^2\left(C_s^2+V_A^2\right)\right]^2+4 k^2 V_A^2\omega_{J}^2} \right.\\
	\left. +\omega_{J0}^2-k^2\left(C_s^2+V_A^2\right)\right],
\end{multline} 
which  at the  Jeans critical wave number $k_{J_2}$ can be estimated as $\gamma\sim\left(k_{J_2}V_A\omega_J\right)^{1/2}$.   
On the other hand, for  $\omega_{J} \lesssim2 \Omega_{0}$, we recast the   dispersion relation \eqref{eq-disp-lamb02}     as 
\begin{equation}
	\label{eq-disp-lamb05}
	\left(\omega^2\right)^2-\left(\omega_{J0}^2+C_S^2 k^2+V_{A}^2k^2\right) \omega^2-\omega_{J}^2 V_{A}^2 k^2=0.
\end{equation} 
Similar to Eq. \eqref{eq-disp-lamb03}, this equation gives also fast and slow wave modes, and it can be shown that while the fast mode is always stable, the slow mode becomes unstable  for any value of $k$. In this case, the growth rate of instability is
\begin{multline}
	\gamma=\frac{1}{\sqrt{2}}\left[\sqrt{\left[\omega_{J0}^2+k^2\left(C_s^2+V_A^2\right)\right]^2+4 k^2 V_A^2\omega_{J}^2} \right.\\
	\left. -\omega_{J0}^2-k^2\left(C_s^2+V_A^2\right)\right],
\end{multline} 
which at the Jeans critical wave number $k_{J_3}$ can be estimated as $\gamma\sim (1/\sqrt{2})k_{J_3}V_A\omega_J/\omega_{J0}\lesssim\omega_J$.
 \par 
Next, in a more general way and in presence of the cosmic rays diffusion, i.e., when $\nu_c\neq0$, we can also similarly discuss the characteristics of different kinds of wave modes including those in the particular cases of  $\lambda=0$ and $\lambda=\pi/2$. However, instead of the classical Jeans mode as in Eq. \eqref{eq-J1}, we can  have a modified Jeans mode, given by,
\begin{equation}
		\omega^2-\tilde{C_s^2} k^2 +\omega_J^2=0, \label{eq-J2}
\end{equation}
 which may exhibit both the damping and the instability (due to the effects of $\nu_c$) similar to those displayed in Fig. \ref{fig2}. Also, in addition to the   fast and slow Alv{\'e}n waves [\textit{cf}. Eq. \eqref{eq-alfven}] there may appear a new wave mode in the intermediate frequency range, i.e., in between the frequencies  of the modified  slow and fast Alfv{\'e}n modes. In the following subsection \ref{sec-sub3}, we will discuss these features in a more general situation, i.e., when the  wave propagation direction and  the  axis of rotation are oblique with respect to the external  magnetic field.  
  \subsection{Propagation at an arbitrary direction} \label{sec-sub3}
In this section, we study the characteristics of MHD waves in a more general situation, i.e., when the direction of propagation $\theta$ is not necessarily $0$ or $\pi/2$, but may assume any value in the interval $0\leq\theta\leq\pi/2$, and the angle $\lambda$ between the  axis of rotation and the $y$-axis is also arbitrary in the interval $0\leq\lambda\leq\pi/2$. Furthermore, we assume that the diffusion frequency $\nu_c$ is not necessarily too small but may be of moderate value, i.e., $\nu_c\lesssim\omega$.
\par
\begin{figure*}
	\includegraphics[width=7in,height=3in]{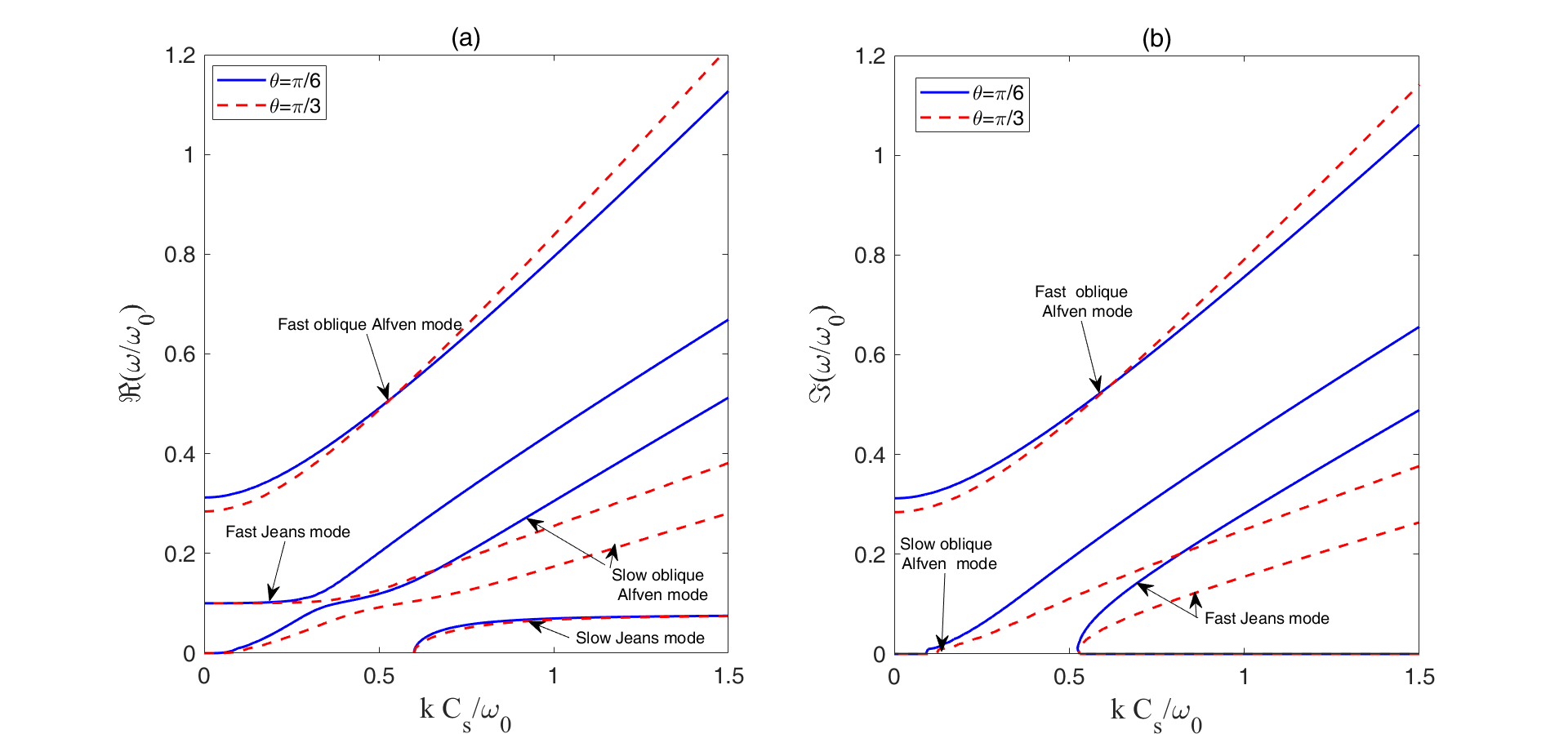}
	\caption{The dispersion curves [subplot (a)] and the instability growth rates [subplot (b)]  of obliquely propagating MHD waves [Eq. \eqref{eq-obliq}] are shown for different values of the angle of propagation $\theta$ as in the legends. The fixed parameter values are $(C_g,~C_c,~V_A)=(0.5,~0.4,~0.5)C_s$ and  $(\nu_c,~\Omega_0,~\omega_J)=(0.1,~0.2,~0.3)\omega_{0}$.  }
	\label{fig3}
\end{figure*}

 
Figure \ref{fig3} shows the dispersion curves [subplot (a)] and the instability growth rates [subplot (b)] of obliquely propagating MHD waves for two different values of   $\theta$ but with a fixed $\lambda=\pi/6$. We find that apart from the fast and slow oblique Alfv{\'e}n modes, there also appear slow and fast Jeans modes due to the effects of the cosmic-ray diffusion. While  the slow Jeans mode appears to be stable with zero growth rate, the fast Jeans mode together with the fast and slow  Alfv{\'e}n modes are, however, unstable. Moreover, such  fast Jeans mode, not reported before in the literature,  appears in the intermediate frequency range, i.e., in between the frequencies of the slow and fast  Alfv{\'e}n waves.  It is also noted that the frequencies of both the fast and slow Alfv{\'e}n waves increase with increasing values of $k$. However, the frequency of the fast Jeans mode initially appears to be more or less a constant for smaller values of $k$ and then starts increasing with $k$. In contrast, the frequency of the slow Jeans mode initially increases and then reaches a steady state at higher values of $k$. For all these modes except the slow Jeans mode, the growth rate of instability has increasing behaviors with increasing values of $k$. Thus, it is noted that the obliqueness of the wave propagation significantly alters the dispersion characteristics as well as the instability growth rates, which can be controlled by suitably tuning the direction of propagation. 
\par
\begin{figure*}
	\includegraphics[width=7in,height=3in]{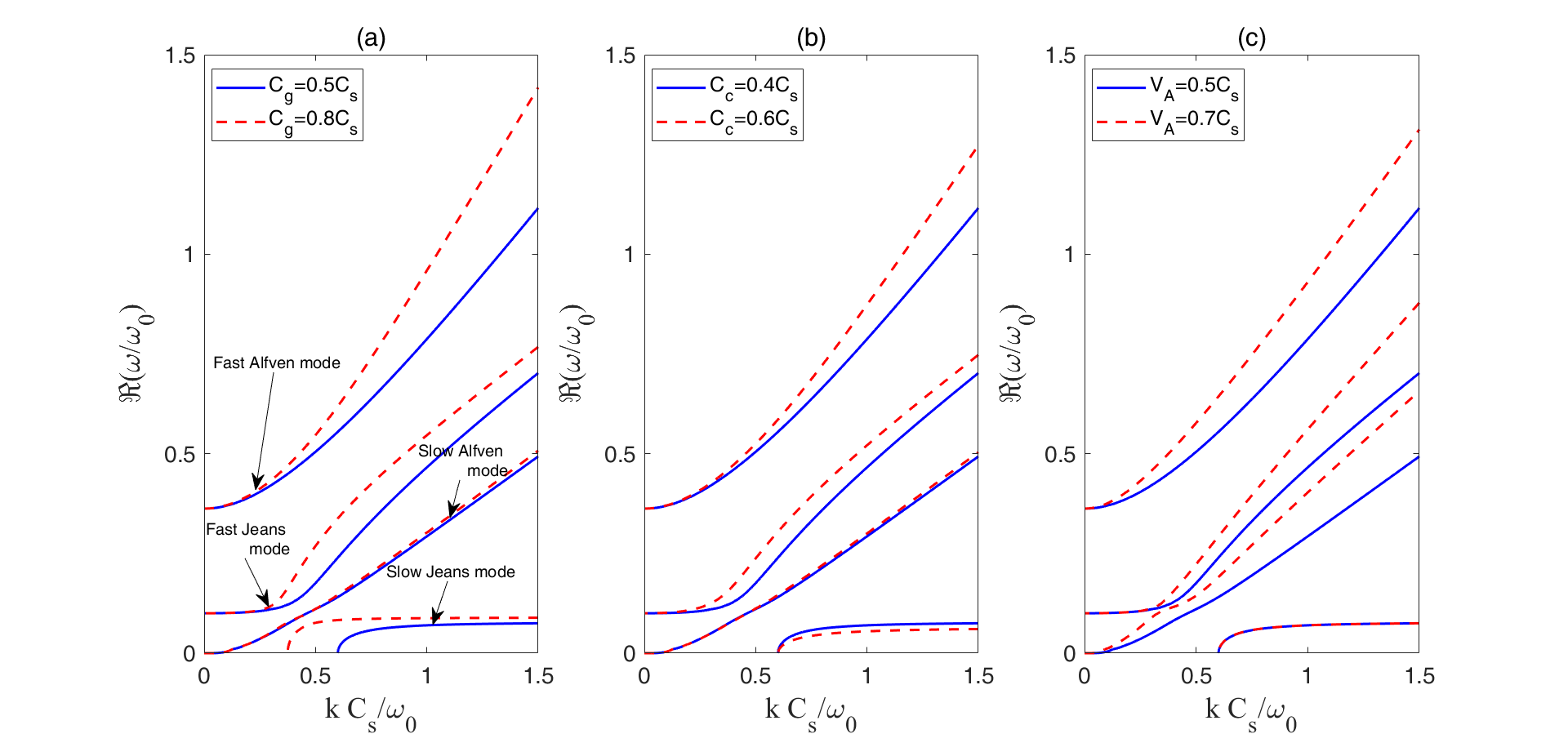}
	\includegraphics[width=7in,height=3in]{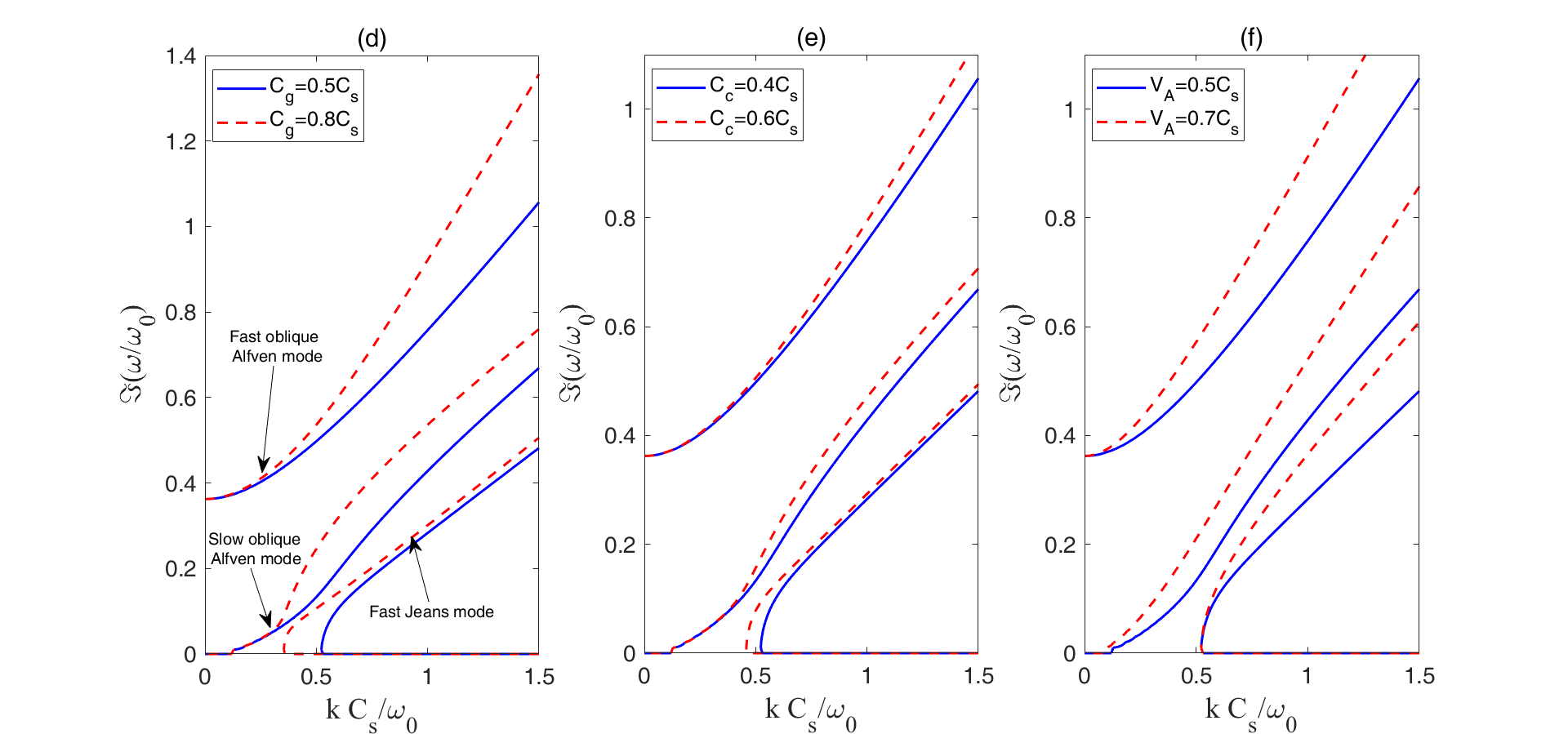}
	\caption{The dispersion curves [subplots (a), (b), and (c)] and the instability growth rates [subplots (d), (e), and (f)]  of obliquely propagating MHD waves [Eq. \eqref{eq-obliq}] are shown for different values of the parameters $C_g$, $C_c$, and $V_A$ as in the legends. The fixed parameter values are   $(\nu_c,~\Omega_0,~\omega_J)=(0.1,~0.2,~0.3)\omega_{0}$. Different wave modes are identified as indicated by text arrows in the subplots (a)  [also applicable to the subplots (b) and (c)] and   (d) [also applicable to the subplots (e) and (f)]. }
	\label{fig4}
\end{figure*}

 
In what follows, we have also studied the characteristics of the wave frequencies and the instability growth rates of  oblique modes by the effects of the parameters that are associated with the gas and cosmic rays pressures ($C_g$ and $C_c$), and  the magnetic field intensity ($V_A$). The results are displayed in Fig. \ref{fig4}. While   subplots (a), (b) and (c) are for the dispersion curves,  the corresponding growth rates are shown in subplots (d), (e) and (f).     From subplots  (a) and (b)   we note that the thermal gas pressure and the cosmic rays pressure have the similar effects on the slow and fast  oblique Alfv{\'e}n waves  in increasing the wave frequencies. However, while the frequency of the slow Jeans mode increases with an increasing value of $C_g$, it decreases with increasing values of $C_c$. Here, both $C_g$ and $C_c$ do not have significant impacts on the slow oblique Alfv{\'e}n waves. Furthermore, while the parameter $V_A$ has no effect on the slow Jeans mode, it increases the frequencies of all other modes [See subplot (c)].
On the other hand, subplots (d), (e) and (f)   show that the growth rates for all the modes are increased with increasing values of $C_g$,    $C_c$ and $V_A$   except for the slow Jeans mode which exhibits stabilizing behaviors with zero growth rate. The enhancement of the growth rate is, however,  noticeable by the effects of    $C_g$   and $V_A$.   
\par
\begin{figure*}
	\includegraphics[width=7in,height=3in]{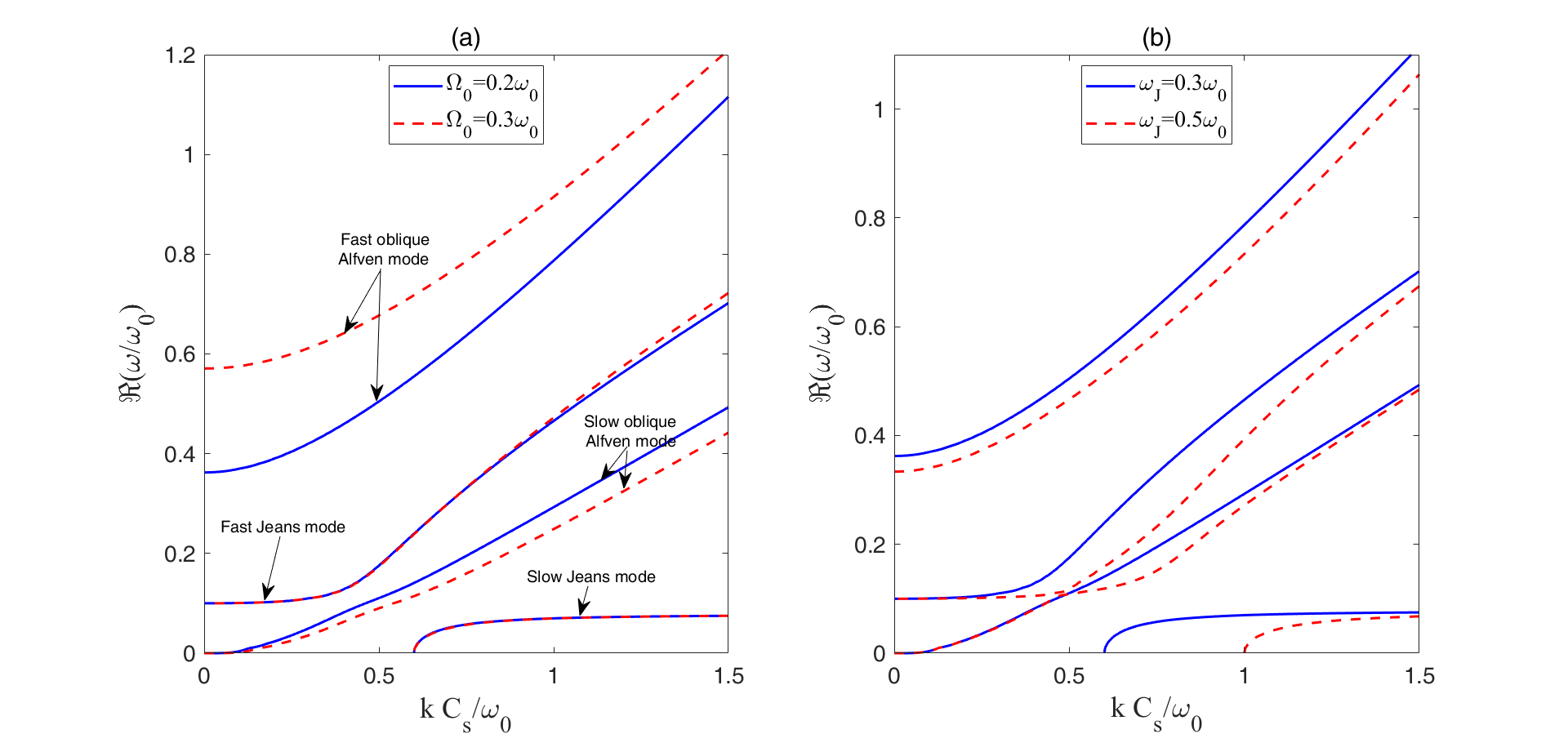}
	\includegraphics[width=7in,height=3in]{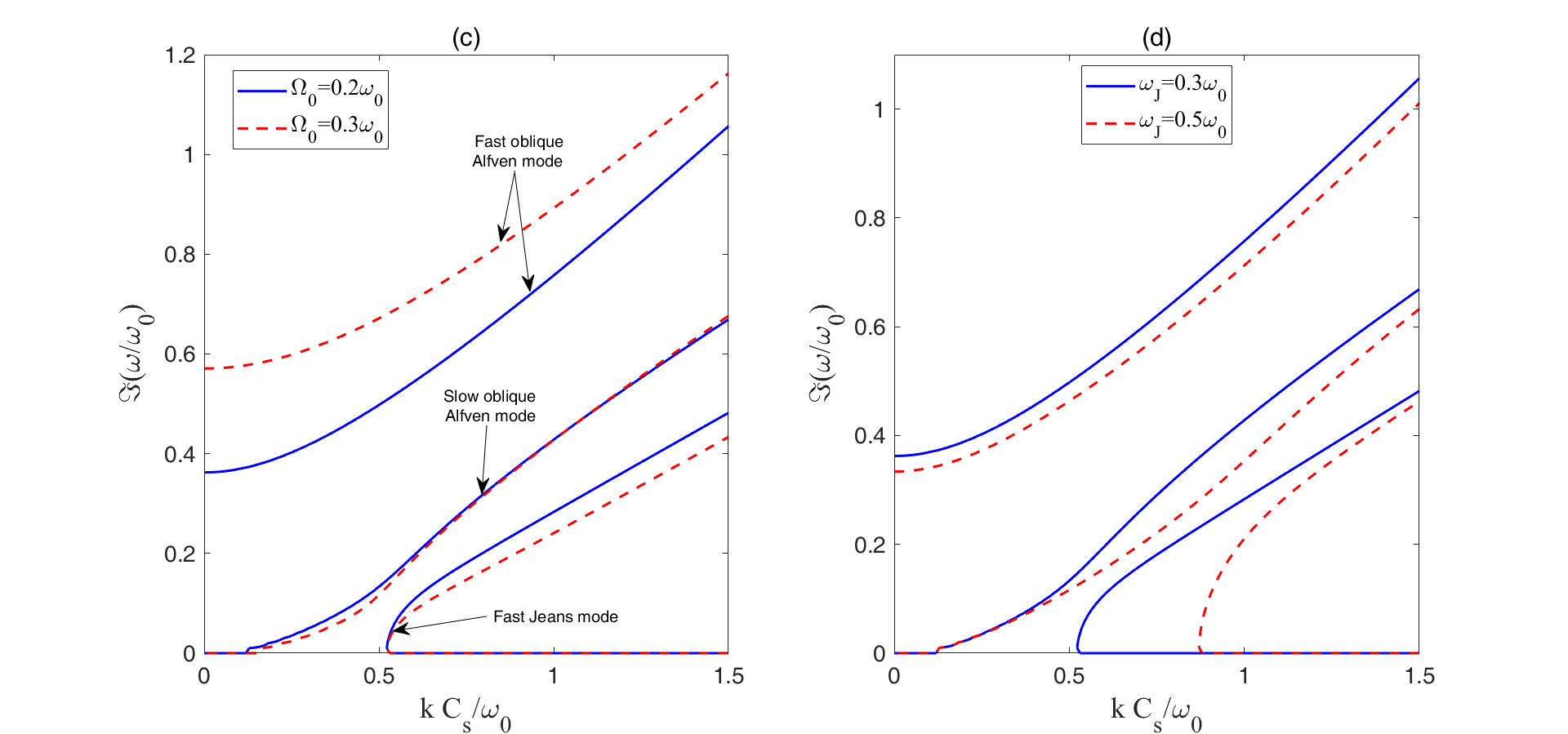}
	\caption{The same as in Fig. \ref{fig4}, but for different values of the parameters $\Omega_0$  and $\omega_J$ as in the legends. The fixed parameter values are $\nu_c=0.1\omega_{0}$ and   $(C_g,~C_c,~V_A)=(0.5,~0.4,~0.5)C_s$. Different wave modes are identified as indicated by text arrows in the subplots (a)  [also applicable to the subplot  (b)] and   (c) [also applicable to the subplots (d)]. }
	\label{fig5}
\end{figure*}

 
It is also instructive to study the influences of the parameters $\Omega_0$ and $\omega_J$ associated with the Coriolis  and  gravitational forces on the dispersion curves and the growth rates of instability of oblique MHD waves which are exhibited in  Fig. \ref{fig5}. As is seen, the  rotational frequency of cosmic fluids does not have any significant influence on the Jeans modes as expected. However, it reduces (increases) the frequency of slow (fast) oblique Alfv{\'e}n modes [subplot (a)]. Subplot (b) shows that the wave frequencies for all the modes are reduced by the effects of the Jeans frequency with its increased values. On the other hand, subplots (c) and (d) show that the effects of $\Omega_0$ are to enhance the growth rate  of the  fast oblique Alfv{\'e}n mode, but to reduce that of the fast Jeans mode. It has no significant effect on the instability of slow Alfv{\'e}n mode. However, the instability growth rates for all the modes (except for the slow Jeans mode) are significantly reduced by the effects of the Jeans frequency. 
   It is also noted that the effect of the cosmic rays diffusion frequency $\nu_c$ has no significant impact on the wave modes and instability growth rates except that it enhances the frequencies of the slow and fast Jeans modes. This happens as from Eq. \eqref{eq-obliq} it is evident that $\nu_c$ contributes only to the factors (with $\tilde{C_s^2}$) associated with the Jeans modes. 
\section{Summary and conclusion}\label{sec-summ}
We have studied the generation of different kinds of MHD waves as well as the characteristics of the wave dispersion and the damping or growth rates of instability in   rotating cosmic magnetoplasmas under the influences of the self-gravitational force, the Coriolis force, both the thermal gas and cosmic rays pressures as well as the cosmic rays diffusion. 
 The main results can be summarized as follows: 
 \begin{itemize}
 \item The Jeans, Alfv{\'e}n and magnetosonic  waves   are shown to be unstable due to the effects of cosmic rays diffusion. They can be damped or anti-damped within the finite domains of wave numbers above or below some critical values. The latter are reduced by the effects of the Coriolis force and the cosmic-ray pressure.
 \item  For the propagation of MHD waves perpendicular to the magnetic field, when the Jeans frequency is less than or comparable to the  rotational frequency, i.e., $\omega_J\lesssim2\Omega_0$,    the   damping
rate of Jeans-Alfv{\'e}n-magnetosonic modes can be reduced by either increasing the thermal gas pressure  (or the magnetic field intensity)  or decreasing the cosmic  rays pressure  (or the Jeans frequency). On the other hand, when  
$\omega_J>2\Omega_0$, both the damping and instability can occur, however, in two different domains of $k$: $k>k_{J_1}$ and $k<k_{J_1}$ respectively, where $k_{J_1}$ is the reduced Jeans critical   wave number. However, the instability prevails if the self-gravity force strongly dominates over the Coriolis force. 
 \item When the propagation direction of MHD waves is parallel to that of the magnetic field, the coupling of Jeans and Alfv{\'e}n waves 
 occurs due to the obliqueness $(0\leq\lambda\leq\pi/2)$ of the  axis of rotation about the  magnetic field. They get decoupled when $\mathbf{\Omega}\parallel\mathbf{B_0}$, i.e., $\lambda=\pi/2$. In the latter, when the effects of the cosmic rays diffusion is ignored, the classical Jeans mode   (modified by the cosmic rays pressure) as well as the fast and slow Alfv{\'e}n modes (modified by the Coriolis force) are recovered. The  fast (slow) Jeans-Alfv{\'e}n mode is, however,    stable (unstable) when $\lambda=0$.
\item In  a more general situation with the effects of the cosmic rays pressure and diffusion, when the propagation direction and the orientation of the  axis of rotation are arbitrary, a new MHD mode, to be called fast Jeans mode, is found to exist for $0\leq\theta<\pi/2$ and $0\leq\lambda\leq\pi/2$   in the intermediate frequency ranges of the slow and fast oblique    Alfv{\'e}n waves. While the slow Jeans mode exhibits stabilizing behaviors, the other modes are always unstable. 
 \end{itemize}
 \par
To conclude, it is to be noted that the insentropic plasma equilibrium specified by the  density, pressure and magnetic field inhomogenities may alter the propagation characteristics of MHD waves significantly, especially in astrophysical environments. However, we have limited our model to those regimes where the gradient length scale  of the plasma equilibrium is much larger than the wavelengths of perturbations. In these regimes, it is possible to identify the MHD wave modes as distinct modes (i.e., Alfv{\'e}n, magnetosonic, fast and slow modes). However, in the case of an inhomogeneous background, the MHD modes may not necessarily be distinct.  We plan to include this effect of plasma inhomogeneity  on the propagation of MHD waves in our future projects.  Furthermore,    apart from the wave damping  of MHD waves to occur, the growth rate of instability can be  typically large in   highly magnetized plasmas with strong gravitational effects. Such high growth rates can be sufficient for the seed perturbations to reach at the nonlinear regimes, thereby contributing to  the gravitational collapse or the formation of large-scale clouds such as those in spiral arms or galactic centers.  
\section*{Acknowledgements}
 The authors would like to thank the anonymous Referees for their valuable comments and suggestions towards improving the manuscript.  One of us, J. Turi wishes to thank Council of Scientific and Industrial Research (CSIR) for a Junior Research Fellowship (JRF) with reference number   09/202(0115)/2020-EMR-I. 
 \section*{Data availability statement}
The data that support the findings of this study are available upon reasonable request from the authors. 
\bibliographystyle{apsrev4-1}
\bibliography{ref}

\end{document}